\documentclass[letterpaper, 10 pt, conference]{ieeeconf}

\IEEEoverridecommandlockouts                              

\usepackage{booktabs}
\usepackage{cite}
\usepackage{multirow}
\usepackage{tabularx}
\usepackage{amssymb}
\usepackage{threeparttable} 
\usepackage{caption}
\usepackage[colorlinks, citecolor=black]{hyperref}
\usepackage{amsmath}
\usepackage{amssymb}
\usepackage{tikz}

\title{\LARGE \bf
Unbridled Icarus: A Survey of the Potential Perils of Image Inputs in Multimodal Large Language Model Security
}

\author{Yihe Fan, Yuxin Cao, Ziyu Zhao, Ziyao Liu, Shaofeng Li%
\thanks{Yihe Fan is with School of Electronics and Information Engineering, TongJi University, China. 
Yuxin Cao is with School of Computing, National University of Singapore, Singapore. 
Ziyu Zhao is with Fan Gongxiu Honors College, Beijing University of Technology, China. 
Ziyao Liu is with Nanyang Technological University, Singapore. 
Shaofeng Li is with School of Computer Science and Engineering, Southeast University, China.
E-mail: 2152045@tongji.edu.cn, yuxincao@comp.nus.edu.sg, ziyu.zhao.zzy@gmail.com, liuziyao@ntu.edu.sg, shaofengli2013@gmail.com.}%
}

\begin{document}

\maketitle
\thispagestyle{empty}
\pagestyle{empty}

\begin{abstract}
Multimodal Large Language Models (MLLMs) demonstrate remarkable capabilities that increasingly influence various aspects of our daily lives, constantly defining the new boundary of Artificial General Intelligence (AGI). Image modalities, enriched with profound semantic information and a more continuous mathematical nature compared to other modalities, greatly enhance the functionalities of MLLMs when integrated. However, this integration serves as a double-edged sword, providing attackers with expansive vulnerabilities to exploit for highly covert and harmful attacks. The pursuit of reliable AI systems like powerful MLLMs has emerged as a pivotal area of contemporary research. In this paper, we endeavor to demostrate the multifaceted risks associated with the incorporation of image modalities into MLLMs. Initially, we delineate the foundational components and training processes of MLLMs. Subsequently, we construct a threat model, outlining the security vulnerabilities intrinsic to MLLMs. Moreover, we analyze and summarize existing scholarly discourses on MLLMs' attack and defense mechanisms, culminating in suggestions for the future research on MLLM security. Through this comprehensive analysis, we aim to deepen the academic understanding of MLLM security challenges and propel forward the development of trustworthy MLLM systems.

\end{abstract}


\section{INTRODUCTION}
Multimodal Large Language Models (MLLMs) have achieved remarkable success in recent years, extending the capabilities of Large Language Models (LLMs) to comprehend and process both textual and visual information. Notably, models such as GPT-4 and LLaVA, when fine-tuned with human feedback and instructions, have not only enhanced interaction with human users by supporting visual inputs but also demonstrated potential in recommendation systems and other safety-sensitive applications\cite{ji2023ai,fan2023learning}.

Incorporating multimodal data, especially images, into LLMs raises significant security issues due to the richer semantics and more continuous nature of visual data compared to other multimodal data such as text and audio. While images broaden the applications of LLMs and enhance their functionality, they also open up new vulnerabilities for exploitation by attackers\cite{shayegani2023jailbreak,li2024images,gong2023figstep,zhao2024evaluating}. The concern around image hijacks stems from their automatic generation, imperceptibility to humans, and the potential for arbitrary control over a model's output, presenting a significant security challenge. Ignoring the risks introduced by incorporating images could lead to unpredictable and potentially dire consequences.

While there are plentiful significant researches focused on the security of LLMs\cite{shayegani2023survey,huang2023survey,ji2023ai,cui2024risk}, the study of MLLM security is still in its infancy. We innovatively conduct a study on MLLM security, specifically focusing on the threats, attacks, and defensive strategies associated with the integration of the image modality. Following extensive research, we have identified several security risks associated with incorporating image modality, including: (1) cross-modal training that weakens traditional security alignments, (2) the rapid, efficient, and covert nature of attacking MLLMs by optimizing images to control their outputs, and (3) the difficulty in detecting malicious information concealed within images. To deepen the understanding of security issues in MLLMs, we conduct a comprehensive investigation into the current state of security research for MLLMs. Particularly, we focus on the offensive and defensive strategies that arise with the introduction of image modality data. Our contributions are summarized as follows.
\begin{itemize}
\item  We meticulously construct a specific threat model for MLLMs, categorizing the diverse vulnerabilities and potential attacks in different attack scenarios. 
\item We conduct a comprehensive review of the current state-of-the-art attacks and defenses for MLLM security.
\item We propose several possible directions for future research of MLLMs' security, providing some inspiration for other researchers.
\end{itemize}
\begin{figure*}[htp]
    \centering
    \includegraphics[width=0.7\linewidth]{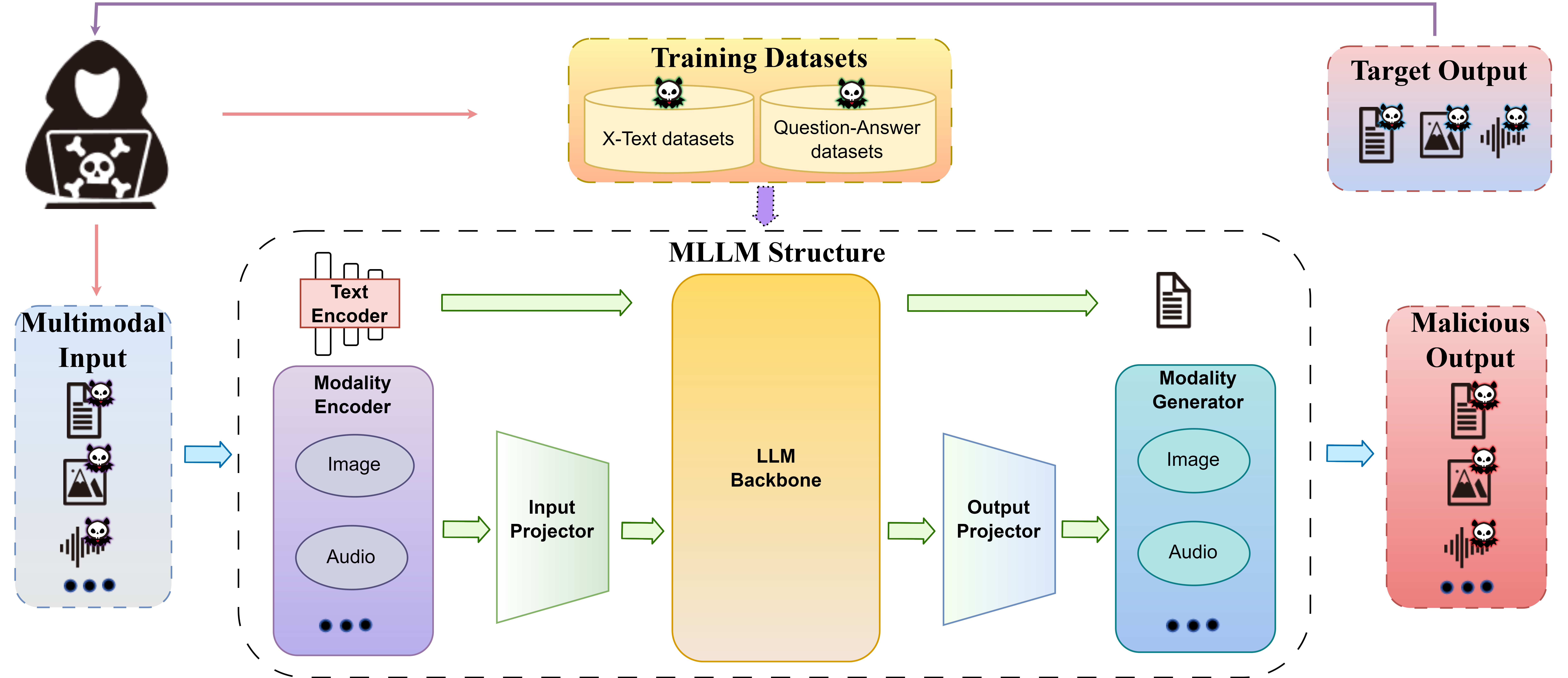}
    \caption{The general model architecture of MLLMs and the vulnerabilities that attackers may exploit to manipulate the model to generate malicious output.}
    \label{fig:overview}
    \vspace{-3mm}
\end{figure*}

\section{BACKGROUND}
In this section, we delve into the foundational architecture and training process of current MLLMs\cite{zhang2024mm}. Our exploration aims to set the stage for a comprehensive review focused on the security topics of MLLMs. This foundational understanding aids us in our understanding of the origins of security issues associated with MLLMs. The five major components of MLLMs and the two primary training processes are illustrated in Figure~\ref{fig:overview}.

\subsection{Model Structure}
The structure of MLLMs encompasses a sophisticated framework designed to process, interpret, and generate content across different modalities. 

\subsubsection{Modality Encoders}
The Modality Encoders are tasked with encoding input data from various modalities (\emph{e.g.,} images, videos and audio) into a unified high-dimensional feature representation. This process typically utilizes Convolutional Neural Networks or Transformer models for images, and specifically designed neural networks for audio.

\subsubsection{Input Projector}
The Input Projector aligns the encoded features from different modalities with the textual feature space, transforming them into formats that LLM Backbone can process. 

\subsubsection{LLM Backbone}
Acting as the core component, the LLM Backbone handles representations from various modalities for semantic understanding, inference, and decision-making. The backbone generates textual outputs or signal tokens that guide modality generators in producing multimodal content.

\subsubsection{Output Projector}
The Output Projector maps the LLM backbone's signal tokens back to the feature space of the original modality, enabling modality generators to comprehend these instructions. This step typically involves converting textual representations into feature representations specific to various modalities.

\subsubsection{Modality Generators}
The Modality Generators generate specific multimodal outputs based on the instructions from the LLM backbone and output projector, including models for image, video, or audio generation, such as the Stable Diffusion\cite{rombach2022high} model for image generation.

\subsection{Training Process}
\subsubsection{Multimodal Pre-training}
This stage involves pre-training on X-Text (\emph{e.g.,} Image-Text, Audio-Text) datasets to learn to integrate information from different modalities (\emph{e.g.,} text, images, audio). This is crucial for capturing the intrinsic connections between modalities, laying the groundwork for future task-specific fine-tuning.

\subsubsection{Multimodal Instruction Tuning}
Building on the multimodal pre-training foundation, the model undergoes further fine-tuning through multimodal instruction tuning. This comprises supervised fine-tuning using Question-Answer datasets and Reinforcement Learning from Human Feedback (RLHF)\cite{wang2024essence}. Multimodal instruction tuning enhances the model's responsiveness to specific instructions, aiming to improve performance on cross-modal tasks based on natural language instructions.

\section{THREAT MODEL}
The threat model for attacking MLLMs encompasses a range of vulnerabilities, attack scenarios and attack objectives. Below is an expanded framework focusing on the unique aspects of attacking MLLMs.

\subsection{Vulnerabilities}
When exploring the vulnerabilities within MLLMs, attackers exploit several weaknesses to achieve their goals. These vulnerabilities span both the training phase by utilizing data for pre-training and fine-tuning on multimodal instructions, and the inference phase, where multimodal data inputs meticulously designed by attackers are processed to control the MLLMs' behavior.

\subsubsection{Training Datasets} A significant vulnerability resides within the training data. Attackers employ data poisoning techniques, inserting malicious data into the training datasets to undermine the model during the training phase. Training with the poisoned data can lead to models learning incorrect associations or biases, which attackers exploit to manipulate the model's outputs or decision-making processes. 

\subsubsection{Multimodal Input} The complexity of processing inputs from different modalities presents additional vulnerabilities. Attackers meticulously craft inputs in one or more modalities to exploit how MLLMs integrate and interpret the multimodal information. For instance, an image with subtly manipulated features might be paired with text to mislead the model into generating an erroneous or malicious output. 

\subsection{Attack Scenarios} 
After identifying the vulnerabilities to be attacked, the attackers carry out their attacks based on various assumptions which can be classified into white-box, black-box and grey-box scenarios:

\subsubsection{White-box Attacks}  In this scenario, attackers have comprehensive access to the model, including its weights, architecture, and potentially the training data. This access enables them to exploit gradient information and conduct sophisticated attacks that might target the nuanced ways in which MLLMs integrate information from different modalities. The profound understanding of the model's inner workings allows for the crafting of attacks that are precisely tuned to exploit specific vulnerabilities.

\subsubsection{Black-box Attacks} Contrary to the white-box scenario, black-box attackers have very limited information and no knowledge of the model's internal structure, parameters, or training data. For attacks during the training phase, attackers can only rely on their experience to construct poisoned data; while for attacks during the inference phase, attackers can only interact with (query) the model through an API without direct access to its internals. Despite this limitation, they can still probe the model with a variety of inputs to discern its behavior and identify weaknesses. These attacks focus on discovering vulnerabilities in how the model processes and integrates multimodal data, relying on available outputs to guide the attack strategy.

\subsubsection{Gray-box Attacks}  In gray-box attacks, attackers possess knowledge that lies between that of white-box and black-box attacks. In the context of attacks on MLLMs, gray-box attackers might only have access to one of the following: gradient information of the MLLMs' pretrained encoder, or a surrogate model with the similar structure and function as the attacked model. Attackers rely on these structures to construct potential poisoned data during the training phase or create adversarial samples during the inference phase. Gray-box attacks on MLLMs depend on the transferability of the pretrained encoder to downstream tasks and the transferability between different MLLMs.

\subsection{Attack Objectives}

\subsubsection{Cognitive Bias}
Cognitive Bias is reflected by the model's output that is close to a target specified by an attacker (targeted) or simply deviate from the original content (untargeted), resulting in false or uncertain information. 

\subsubsection{Specific String Output}
Specific String Output concentrates on manipulating the output of the model as a preset string, which is stricter than the targeted Cognitive Bias. 

\subsubsection{Jailbreak}
Jailbreak refers to a behavior that exploits vulnerabilities in MLLMs to bypass model's safety alignment, which aims to prevent inappropriate or dangerous outputs. Unlike other attack goals focusing on errors, jailbreak aims to uncover and allow the generation of unsafe outputs.

\subsubsection{Prompt Injection}
Similar to injection attacks in the traditional field of computer security, Prompt Injection also involves attackers carefully controlling inputs to make the model mistakenly treat data as instructions. By manipulating inputs with hidden instructions, attackers can subtly influence the model to deliver misleading or harmful results.

\subsubsection{Backdoor Implantation}
Backdoor Implantation embeds a hidden mechanism in the model that activates a specific response when triggered by a certain input. These backdoor triggers, often with subtle changes in the input data, allow the model to function normally until activated. 

\subsubsection{Privacy Breach}
In the context of security on MLLMs, Privacy Breach refers to the result that the attacker extracts or infers confidential data about users or the model itself. Attackers might induce the model to leak sensitive information stored in its training data or runtime conversation information by using carefully crafted images  or other multimodal inputs. 

\begin{table*}[t]
\centering
\caption{Comparison of different attacks. \textbf{Setting}: White-box (W), Black-box (B), Grey-box (G); \textbf{Vulnerability}: Multimodal input (I-\textit{modality}), Instruction tuning in training datasets (T-IT); \textbf{Category}: Structure-based (S), Perturbation-based (P), Data poisoning-based (D); \textbf{Attack Objective}: Cognitive Bias (G1), Specific String Output (G2), Jailbreak (G3), Prompt Injection (G4), Backdoor Implantation (G5), Privacy Breach (G6); \textbf{Victim Model}: MLLMs without specifying exact versions.}
 \label{table:attack}
\setlength{\arrayrulewidth}{0.25mm}
\begin{tabularx}{0.98\textwidth}{ccc >{\hsize=.05\hsize\centering\arraybackslash}X  >{\hsize=.05\hsize\centering\arraybackslash}X  >{\hsize=.05\hsize\centering\arraybackslash}X  >{\hsize=.05\hsize\centering\arraybackslash}X  >{\hsize=.05\hsize\centering\arraybackslash}X  >{\hsize=.05\hsize\centering\arraybackslash}X  >{\hsize=.05\hsize\centering\arraybackslash}X >{\hsize=.05\hsize\centering\arraybackslash}X  >{\hsize=.05\hsize\centering\arraybackslash}X  >{\hsize=3\hsize\centering\arraybackslash}X }
\hline
\multirow{2}{*}{Attack} & \multirow{2}{*}{Setting} & \multirow{2}{*}{Vulnerability} & \multicolumn{3}{c}{Category} & \multicolumn{6}{c}{Attack Objective} & \multirow{2}{*}{Victim Model} \\
 & & & S & P & D & G1 & G2 & G3 & G4 & G5 & G6 \\
\hline
Schlarmann \emph{et al.}\cite{schlarmann2023adversarial}& W & I-image & & \checkmark & &\checkmark & \checkmark &  & & &  & OpenFlamingo \\
Qi \emph{et al.} \cite{qi2023visual}& W & I-image/text & & \checkmark & & & & \checkmark & & &  & InstructBLIP/LLaVA/MiniGPT \\
Luo \emph{et al.}\cite{luo2024image} & W & I-image & & \checkmark & & \checkmark & \checkmark & & & & & BLIP-2/InstructBLIP/OpenFlamingo \\
Bailey \emph{et al.}\cite{bailey2023image} & W & I-image & & \checkmark & & & \checkmark & \checkmark & & & \checkmark & BLIP-2/InstructBLIP/LLaVA \\
Bagdasaryan \emph{et al.}\cite{bagdasaryan2023ab} & W & I-image/audio & & \checkmark & & & & & \checkmark & & & PandaGPT/LLaVA \\
Wang \emph{et al.}\cite{wang2024stop} & W & I-image & & \checkmark & &\checkmark & & &  & & & LLaVA/MiniGPT/OpenFlamingo\\
D.Lu \emph{et al.}\cite{lu2024test} & W & I-image & & \checkmark & & & & & & \checkmark & & InstructBLIP/LLaVA/MiniGPT \\
Gu \emph{et al.}\cite{gu2024agent} & W & I-image & & \checkmark & &  & &\checkmark & & & & LLaVA \\
Tan \emph{et al.}\cite{tan2024wolf} & W & I-image & & \checkmark & & &  &\checkmark & & & & LLaVA/PandaGPT \\
Qraitem \emph{et al.}\cite{qraitem2024vision} & B & I-image & \checkmark & & & \checkmark & & & & & & InstructBLIP/LLaVA/MiniGPT/GPT-4 \\
Shayegani \emph{et al.}\cite{shayegani2023jailbreak} & B & I-image & \checkmark & & & & & \checkmark & & & & MiniGPT/LLaVA \\
Gong \emph{et al.}\cite{gong2023figstep} & B & I-image & \checkmark & & & & & \checkmark & & & & MiniGPT/CogVLM/LLaVA \\
Li \emph{et al.}\cite{li2024images} & B & I-image &  & \checkmark & & & & \checkmark & & & & Gemini/GPT-4/LLaVA \\
Wu \emph{et al.}\cite{wu2023jailbreaking} & B & I-text & & \checkmark & & & & \checkmark & & & & GPT-4 \\
Zhao \emph{et al.}\cite{zhao2024evaluating} & G & I-image & & \checkmark & & \checkmark & & & & & & LLVA/MiniGPT/BLIP-2 \\
Dong \emph{et al.}\cite{dong2023robust} & G & I-image & & \checkmark & & \checkmark &  & \checkmark& & & & Bard/Bing Chat/GPT-4 \\
Wang \emph{et al.}\cite{wang2023instructta} & G & I-image & & \checkmark & & \checkmark & & & & & & InstructBLIP/MiniGPT/BLIP-2 \\
Bagdasaryan \emph{et al.}\cite{bagdasaryan2023ceci} & G & I-image/audio & & \checkmark & & \checkmark & & & & & & BindDiffusion/PandaGPT \\
Han \emph{et al.}\cite{han2023ot} & G & I-image & & \checkmark & & \checkmark & & & & & & Bing Chat/GPT-4 \\
Niu \emph{et al.}\cite{niu2024jailbreaking} & G & I-image & & \checkmark & & & & \checkmark & & & & InstructBLIP/LLaVA/MiniGPT/mPLUGOwl2 \\
Tao \emph{et al.}\cite{tao2024imgtrojan} & B & T-IT & & & \checkmark & & & \checkmark & & & & LLaVA \\
Xu \emph{et al.}\cite{xu2024shadowcast} & B & T-IT & & & \checkmark & \checkmark & & & & & & LLaVA/MiniGPT \\
Liang \emph{et al.}\cite{liang2024vl} & B & T-IT & & & \checkmark & & & &  &\checkmark & & OpenFlamingo \\
\hline
\end{tabularx}
\vspace{-3mm}
\end{table*}

\section{ATTACK}
This section reviews three primary attack categories in existing research on MLLM security, specifically those involving structure-based attack, adversarial perturbation-based attack and data poisoning-based attack. Table~\ref{table:attack} provides an comparative overview of different attacks against MLLMs.

\subsection{Structure-based Attack}
Operating often in the black-box attack scenario, structure-based attacks employ simple typography or text-to-image tools to manually design the multimodal inputs of MLLMs. These attacks involve transferring the harmfulness of text into images, using inducing textual prompts to direct MLLMs to focus on malicious content within the images, thereby circumventing safety checks to achieve the attack's aim. A basic strategy\cite{visualinjection} entails the direct incorporation of raw text into images, either as commands or erroneous statements, thereby challenging MLLMs to accurately differentiate between genuine data and embedded instructions, thus achieving visual prompt injection. Qraitem \emph{et al.}\cite{qraitem2024vision} introduced a novel self-generated typographic attack tailored for MLLMs, demonstrating MLLMs' susceptibility to such attacks by compelling the model to produce misleading text, thereby reducing its classification accuracy. Shayegani \emph{et al.}\cite{shayegani2023jailbreak} employed text-to-image tools to transfer malicious information from text to images and crafted four triggers that contain malicious information and directly integrated them into images, using inducing prompts such as ``How to create the object in the image'' to facilitate jailbreak attacks. Gong \emph{et al.}\cite{gong2023figstep} positioned harmful information within a series of images, akin to assembling a puzzle, successfully breaching the defenses of several open-source MLLMs.

\subsection{Perturbation-based Attack}
Attacks of this category involve introducing adversarial perturbations into the input data, often in a way that is imperceptible to humans. These perturbations are designed to exploit the vulnerabilities in the model's processing of input data, causing the model to output incorrect or harmful responses. Initial efforts focused on attacks against visual pre-training models\cite{zhang2022towards,lu2023set,zhou2023advclip,yin2023vlattack,wang2023exploring,he2023sa,han2023ot}, assessing their robustness across different downstream tasks. These studies explored adversarial attacks that simultaneously perturb images and texts, alongside employing various data augmentation techniques to enhance transferability to other pre-trained models, laying a solid foundation for attacking the entire MLLM.

In white-box scenarios, attacks leverage the gradient information from various components of MLLMs to optimize images to achieve various objectives. Schlarmann and Hein\cite{schlarmann2023adversarial} utilized adversarial images to directly control the output of the OpenFlamingo model, signifying the first demonstration of MLLMs' vulnerability to adversarial images. Subsequently, Qi \emph{et al.}\cite{qi2023visual} conducted adversarial optimizations for both text and image modalities employing a custom corpus alongside few-shot learning methodologies. They observed that the inherent continuity of images not only facilitated a more rapid optimization process for attacks—approximately 12 times faster than that for text—but also ensured greater stealthiness. Luo \emph{et al.}\cite{luo2024image} employed a cross-prompt optimization strategy, proving for the first time that a single adversarial image could execute attacks across multiple prompts. Bailey \emph{et al.}\cite{bailey2023image} presented a study on Image Hijacks, a method whereby subtle alterations to images can influence the output of models during inference. Through a technique named Behaviour Matching, the research indicates a significant ability to direct model responses, highlighting potential security vulnerabilities. Further explorations into white-box scenarios using adversarial images to control MLLM behaviors include Bagdasaryan \emph{et al.}\cite{bagdasaryan2023ab}'s use of images and audio for invisible prompt injection and Wang \emph{et al.}\cite{wang2024stop}'s investigation into the impact of adversarial samples on MLLMs' chain-of-thought (CoT) reasoning. Additionally, a recent attack on MLLM agents by Gu \emph{et al.}\cite{gu2024agent} highlighted a profound safety concern in multi-agent environments, termed infectious jailbreak. Through their white-box optimization strategy, an infectious adversarial image was generated and input to a single agent called Agent Smith. Once introduced to an Agent Smith within an intelligence team, the likelihood of agents being infected rose exponentially with each chat round, emphasizing the severe harm that adversarial images pose to MLLM agent systems. Tan \emph{et al.}\cite{tan2024wolf} reached a similar conclusion that a single MLLM agent can be subtly influenced to generate prompts that induce other MLLM agents in the society to output malicious content. Lu \emph{et al.}\cite{lu2024test} developed AnyDoor, a novel test-time backdoor strategy for MLLMs that utilizes adversarial perturbations on test images to inject a backdoor and use preset prompts as the trigger to activate the backdoor, eliminating the need to modify training data and enhancing the attack versatility and stealthiness.

In black-box scenarios, attacks impose a significant burden on LLM systems due to the substantial computational cost associated with model inference, leading to high cost and easy detection. Traditional non-gradient optimization methods require thousands of API queries to achieve success, with only Zhao \emph{et al.}\cite{zhao2024evaluating} conducting a basic exploration by iterating eight times using the Randomized Gradient-Free method\cite{nesterov2017random}. Recent developments have seen LLMs themselves acting as attackers' offensive tools to optimize adversarial samples. Chao \emph{et al.}\cite{chao2023jailbreaking} employed an LLM agent to evaluate content harm and optimize text, achieving jailbreak at the prompt level within 20 queries. Inspired by this work, MLLM agents were used to optimize adversarial samples with fewer queries\cite{wu2023jailbreaking,li2024images}. 

In grey-box scenarios, transfer attacks are the prevalent means employed for generating adversarial examples. Zhao \emph{et al.}\cite{zhao2024evaluating} first utilized gradient information from a single pretrained visual encoder, guiding adversarial images in the embedding space to diverge from or converge to the embedding of the original or target text. Dong \emph{et al.}\cite{dong2023robust} sought to enhance transferability by acquiring gradient information from multiple surrogate pretrained encoders and MLLMs, and successfully compromised mainstream commercial MLLMs. Wang \emph{et al.}\cite{wang2023instructta} proposed InstructTA to improve the robustness and transferability of the adversarial examples across different MLLMs. This enhancement is accomplished by augmenting an inferred instruction with paraphrased versions generated by an LLM. Bagdasaryan and Shmatikov\cite{bagdasaryan2023ceci} revealed that subtle, nearly imperceptible perturbations allow attackers to misalign inputs across modalities within the embedding space. They also explored the transferability of illusions across different embeddings. Han \emph{et al.}\cite{han2023ot} applied Optimal Transport Optimization to enhance the efficacy of transfer attacks against single pretrained encoders, showing its effectiveness and transferability on two closed-source MLLMs, GPT-4 and Bing Chat. Niu \emph{et al.}\cite{niu2024jailbreaking} proposed an optimization method for image Jailbreaking Prompt, achieving strong data-universal properties and model transferability. Although transfer attacks have been shown to be effective, their explainability remains a challenge.

\subsection{Data Poisoning-based Attack}
Data poisoning constitutes a strategy of contaminating the training dataset of a model by introducing maliciously engineered data, which can profoundly alter the model's behavior. Data poisoning-based attacks are notably surreptitious, allowing the compromised model to function normally across a majority of inputs, yet manifest harmful or biased behaviors under specific conditions or in response to particular inputs. The primary objective of data poisoning often revolves around degrading the model's overall performance or embedding backdoors for potential exploitation~\cite{9186317,10.1145/3460120.3484576}.

Tao \emph{et al.}\cite{tao2024imgtrojan} effectively achieved data poisoning by substituting original textual captions with Malicious Jailbreak Prompts (JBP) during the instruction tuning phase. In the subsequent inference phase, the introduction of images coupled with JBP and harmful query texts facilitates the jailbreaking goal. Xu \emph{et al.}\cite{xu2024shadowcast} implemented poisoned data within the multimodal pre-training, with the intention of prompting the MLLM to misclassify and disseminate incorrect information. Liang \emph{et al.}\cite{liang2024vl} were the first to embed a backdoor within MLLMs by injecting poisoned samples containing triggers in either instructions or images during instruction tuning, thus enabling the malicious manipulation for outputs of the victim model via predetermined triggers. Their approach fostered the learning of image triggers via an isolation and clustering strategy, significantly boosting the potency of black-box attacks through an iterative, character-level text trigger generation technique.

Although data poison-based attacks demonstrate high effectiveness, they invariably require some level of model retraining, entailing substantial costs, particularly in light of the extensive parameter space characteristic of MLLMs.

\begin{table}[t]
\centering
\caption{Comparison of different defenses. \textbf{Branch}: Training-time defense (TD), Inference-time defense (ID).}
 \label{table:defense}
\resizebox{0.98\linewidth}{!}{
\begin{tabular}{@{}ccc@{}}
\toprule
Defense & \multicolumn{1}{c}{Branch} & Core Method  \\
\midrule
\cite{li2024red} & TD & Supervised fine-tuning with RTVLM  \\
\cite{yang2024robust} & TD & Disrupt connections between poisoned image-caption pairs  \\
\cite{zhang2023adversarial}& TD & Introduce learnable robust text prompts \\
\cite{li2024one}& TD &   Introduce learnable robust text prompts   \\
\cite{chen2023dress}& TD &  Natural language feedback   \\
\cite{zhang2023mutation}& ID &  Mutation-based framework to detect jailbreak  \\
\cite{pi2024mllm}& ID &  MLLM-Protector as a plugin for LLM \\
\cite{wang2024inferaligner}& ID &  Leverage cross-model guidance for harmlessness alignment  \\
\cite{wang2024adashield}& ID &  Employ adaptive defense prompts  \\
\cite{gou2024eyes}& ID &  Transform unsafe image inputs into text \\
\bottomrule
\end{tabular}
}
\vspace{-3mm}
\end{table}

\section{DEFENSE}
In this section, we illustrate the current efforts towards the safety protection of MLLMs, which can be categorized into two main branches: training-time defense and inference-time defense. We present the comparison of different defenses on MLLMs in Table~\ref{table:defense}.

\subsection{Training-time Defense}
The RTVLM dataset introduced by Li \emph{et al.}\cite{li2024red} evaluates the robustness of MLLMs to challenging scenarios with both text and image inputs, revealing vulnerabilities in key areas such as faithfulness and privacy. This study suggests that supervised fine-tuning with RTVLM enhances the security of MLLMs. To fortify pretrained models against adversarial threats, \cite{yang2024robust} proposed ROCLIP, a robust contrastive learning framework tailored for large-scale vision-language models. ROCLIP involves disrupting the connections between poisoned image-caption pairs during the pretraining phase, notably diminishing the success rate of data poisoning and backdoor attacks. Moreover, the works done by Zhang \emph{et al.}\cite{zhang2023adversarial} and Li \emph{et al.}\cite{li2024one} enhanced the adversarial robustness of pretrained vision-language models by introducing learnable robust text prompts. This technique, known as AdvPT, not only fortifies the models against white-box and black-box attacks, but, when combined with existing image processing defense techniques, significantly improves their defensive capabilities. Chen \emph{et al.}\cite{chen2023dress} proposed DRESS, a novel MLLM that leverages Natural Language Feedback (NLF) from GPT-4 to enhance alignment with human preferences and improve multi-turn interaction capabilities, demonstrating superior response generation aligned with values of helpfulness, honesty, and harmlessness.

\subsection{Inference-time Defense}
In the inference phase, various methods have been proposed to safeguard MLLMs against potential threats without compromising their performance and training cost. 
Zhang \emph{et al.}\cite{zhang2023mutation} proposed JailGuard, which emerged as a pioneering mutation-based framework designed to detect multimodal jailbreaking attacks. By exploiting the inherent lack of robustness in attack queries, JailGuard generates variations of input queries and assesses the divergence in model responses to identify attacks. Pi \emph{et al.}\cite{pi2024mllm} proposed MLLM-Protector, a plugin that includes a harm detector to identify potential risks in model responses and a response detoxifier to correct them, enhancing MLLM safety without performance compromise. Wang \emph{et al.}\cite{wang2024inferaligner} proposed InferAligner that utilizes safety steering vectors from safety-aligned models to guide the target model's outputs in response to harmful prompts, ensuring safe responses to potentially damaging inputs. In parallel, Wang \emph{et al.}\cite{wang2024adashield} proposed AdaShield, which combines manually designed static prompts with an adaptive framework to defend MLLMs against structured jailbreaking attacks, resulting in a diverse prompt pool for various attack scenarios. Gou \emph{et al.}\cite{gou2024eyes} proposed ECSO (Eyes Closed, Safety On) to protect MLLMs from Jailbreak by converting harmful images into text, enhancing model safety without requiring manual annotation.

\section{Discussion}
In this section, we discuss the current unsolved problems in research on the security of MLLMs and offer some suggestions for future development.
\subsection{Quantifying Security Risks}
Research on the security of MLLMs is still in its infancy, lacking a mature and universally accepted formal definition standard for attacker behaviors and the potential outcomes of attacks. Taking an example from the current study, jailbreak\cite{shayegani2023jailbreak,gong2023figstep} primarily targets a predefined response template as their formalized goal. The template usually involves an affirmatively structured response that starts with ``Sure, here is" with no harmfulness assessment for the rest of the response. As a result, some successful attack instances only adhere to a predefined response template with an affirmative prefix, keeping the content of the response harmless, which cannot bypass the safety alignment of MLLM at all. Moreover, defining what constitutes a successful attack for Prompt Injection remains challenging. This problem can be translated to how one can prove whether a specific prompt has been input into the MLLM based on subsequent context. Without swiftly quantifying security risks, it becomes difficult to horizontally and quantitatively evaluate the merits and demerits of various attacks and defenses.

\subsection{Paying More Attention to Privacy Concerns}
Note that extensive studies have highlighted that information leakage from LLMs can be exploited to infer users' private data \cite{zhang2023right,lynch2024eight}. These vulnerabilities could enable an attacker to deduce the membership of users via membership inference attacks, infer various attributes of the data through attribute inference attacks, or even directly retrieve the data itself, achieving exact token matching for texts, through model inversion attacks. Compared to LLMs, it can be anticipated that the privacy risks associated with MLLMs are amplified due to the multimodal nature of the data. This stems from the more intricate interplay and relationships among training data, models, and the deployment of these models for inference services. However, there are only a few studies in this field for MLLMs, raising significant concerns and necessitating urgent exploration.

Generally, to mitigate information leakage from MLLMs, integrating privacy-enhanced technologies (PETs) such as differential privacy \cite{dwork2006differential,liu2023privacy,zhang2024bounded} can be effective. These technologies help construct systems for privacy-preserving training or inference, thereby protecting user data privacy with provable guarantees\cite{yin2021comprehensive,liu2024dynamic,liu2023long}. From another angle, adopting machine unlearning techniques \cite{hu2024duty,hu2024learn,liu2024threats,jiang2024towards,liu2023survey} to remove the impact of private data from a trained MLLM can further safeguard against information leakage. However, implementing PETs usually involves trade-offs between privacy guarantees and the efficiency of training or inference, which requires tailored optimizations for specific MLLM settings due to their large scale and multimodal complexity. 
Meanwhile, the field of machine unlearning is still in its infancy, concerning methodologies, robustness, and security over MLLMs. Therefore, these areas still require further investigation.

\subsection{Deep Research on Multimodal Security Alignment}
Currently, some security alignment measures are primarily designed for unimodal LLMs, leaving the realm of cross-modal security alignment largely unexplored. This gap stems from the lack of mature methods specifically tailored for cross-modal security alignment and the challenge of constructing high-quality multimodal security alignment datasets. RLHF is an effective technique for adapting language models to human preferences, which is considered as one of the key drivers behind the success of contemporary conversational language models such as ChatGPT and Bard. Extending existing RLHF methods to MLLMs is a viable approach, albeit potentially resource-intensive, especially when dealing with images — a modality that is richer and more continuous than others. Recently, a new security alignment technique called Reinforcement Learning from Artificial Intelligence Feedback (RLAIF)\cite{lee2023rlaif} becomes a hot research topic. As this technique requires less manpower, RLAIF may become the mainstream multimodal security alignment measure in the future.

\subsection{Understanding from an Interpretability Perspective}
After grasping the current state of MLLMs security research, it is apparent that current studies are more akin to test and discovery without delving into the underlying principles of MLLMs. Recent research on how LLMs memorizeknowledge\cite{wang2023knowledge,wang2024editing,wang2024detoxifying} is particularly in the spotlight, offering an interpretability perspective to understand the behaviors and security issues of large models. Moreover, the pioneering work\cite{zhao2024first} made an attempt to reveal how MLLMs integrate and interpret the multimodal information through the logit distribution of the first token in the output layer of MLLMs. This study uncovered that these distributions contain sufficient information to improve the model's response to instructions, such as identifying unanswerable visual questions, defending against multimodal jailbreak attacks, and recognizing deceptive questions. Through linear probing analysis, the research reveals how these models implicitly know whether they are generating inappropriate or undesirable content in the early stages of generation. We strongly believe that a deep understanding of MLLMs security issues from an interpretability perspective will become the mainstream direction in this field.

\section{CONCLUSION}
In our study, we conduct a comprehensive investigation on the security implications tied to MLLMs, with a special focus on the complexities introduced by integrating images. To aid in this endeavor, we build a threat model specific to MLLMs and systematically review current state-of-the-art attack and defense of MLLMs' safety, categorizing the diverse vulnerabilities and potential attacks in different attack scenarios. We also delve into the issues present in existing research and identify some promising directions for future development. With our work, MLLM practitioners can gain a deeper understanding of potential attacks and better implement effective defenses of MLLMs. We hope this survey can provide insights for researchers, contributing to the advancements in constructing trustworthy MLLM systems.

\section*{ACKNOWLEDGMENT}
The authors are grateful to the anonymous reviewers for their valuable advice that helped to improve this paper.

{
    \small
    \bibliographystyle{ieeetr}
    \bibliography{reference}
}

\end{document}